\documentclass[paper,notoc,nohyper]{JHEP3}

\usepackage{amsmath}

\newcommand{\order}[1]{\ensuremath{O\left(#1\right)}}
\newcommand{\hpl}[2]{\ensuremath{H\left(#1;#2\right)}}

\newcommand{\gstar}{\gamma^*}
\newcommand{\en}{E_{cm}}
\newcommand{\enen}{E_{cm}^2}
\newcommand{\enenen}{E_{cm}^4}
\newcommand{\as}{\alpha_s}
\newcommand{\cf}{\ensuremath{C_F}}
\newcommand{\abs}[1]{\left| #1 \right|}
\newcommand{\gaussf}[4]{\ensuremath{\, _2F_1 \left(#1,#2,#3;#4\right)}}

\newcommand{\gammavertex}{\ensuremath{\Gamma^{had}_{\gamma^*, {vertex}}}}
\newcommand{\gammareal}{\ensuremath{\Gamma^{had}_{\gamma^*, \text{real}}}}
\newcommand{\mtwo}{\mathcal{M}^{tree}_{\gstar \rightarrow q \bar{q}}}

\newcommand{\inta}[1]{\mathcal{#1}}
\newcommand{\A}{\ensuremath{A^{0}_{Qg\bar{Q}}}}
\newcommand{\Aint}{\ensuremath{\inta{A}^{0}_{Qg\bar{Q}}}}
\newcommand{\D}{\ensuremath{D^{0}_{Qgg}}}
\newcommand{\Dint}{\ensuremath{\inta{D}^{0}_{Qgg}}}
\newcommand{\Eqm}{\ensuremath{E^{0}_{Qq'\bar{q}'}}}
\newcommand{\Eqmint}{\ensuremath{\inta{E}^{0}_{Qq'\bar{q}'}}}
\newcommand{\Eqprimem}{\ensuremath{E^{0}_{qQ'\bar{Q}'}}}
\newcommand{\Eqprimemint}{\ensuremath{\inta{E}^{0}_{qQ'\bar{Q}'}}}
\newcommand{\G}{\ensuremath{G^{0}_{gQ'\bar{Q}'}}}
\newcommand{\Gint}{\ensuremath{\inta{G}^{0}_{gQ'\bar{Q}'}}}

\newcommand{\splitqgq}{\ensuremath{P_{qg\rightarrow Q}}}
\newcommand{\splitqqbarg}{\ensuremath{P_{q\bar{q}\rightarrow G}}}
\newcommand{\splitggg}{\ensuremath{P_{g g\rightarrow G}}}
\newcommand{\softeikonal}{\ensuremath{\mathcal{S}}}

\newcommand{\form}{FORM}

\newcommand{\fire}{FIRE}

\newcommand{\eps}{\epsilon}
\def\d{\hbox{d}}
\def\JET{J}
\newcommand{\intphimmzero}{\ensuremath{I^{(m,0,m)}_1}}
\newcommand{\intsmmzero}{\ensuremath{I^{(m,0,m)}_2}}
\newcommand{\intphizerozerom}{\ensuremath{I^{(0,0,m)}_1}}
\newcommand{\intszerozerom}{\ensuremath{I^{(0,0,m)}_2}}
\newcommand{\intdiff}{\ensuremath{I^{(0,0,m)}_3}}

\DeclareMathOperator{\li}{Li_2}

\title{\boldmath 
NLO antenna subtraction with massive fermions
}
\author{
A.~Gehrmann--De Ridder, M.~Ritzmann\\
Institute for Theoretical Physics, ETH, CH-8093 Z\"urich,
Switzerland}
\abstract{ 
We present an extension of the antenna subtraction formalism at NLO to include 
massive final state fermions. The basic ingredients to the subtraction terms, 
the NLO massive final-final antenna 
functions are derived and integrated over the corresponding 
factorised phase space. Those antenna functions account 
for all soft, collinear and quasi-collinear limits of the QCD matrix elements 
involving massive fermions in the final state.} 
\begin{document}

\renewcommand{\theequation}{\mbox{\arabic{section}.\arabic{equation}}}

\section{Introduction}
\setcounter{equation}{0}
Hard scattering processes leading to final states 
with heavy particles are important observables for present and future 
colliders. Reliable theoretical predictions for these observables require 
the calculation of at least the next-to-leading order QCD corrections.

For hard scattering processes, 
involving massless or massive QCD partons, 
the perturbative corrections to a given process 
and at a given order in QCD are obtained when  
all partonic channels contributing to that order are summed.
In general, each partonic channel contains both ultraviolet and infrared
(soft and collinear) divergences.
The ultraviolet poles are removed by renormalisation in each channel.
The remaining soft and collinear poles cancel among each other when all  
partonic contributions are summed over \cite{KLN}.      
At NLO, one has to combine virtual-one loop 
calculations with real emission contributions from unresolved partons.
While infrared singularities from purely virtual corrections 
are obtained immediately after integration over the loop momenta, 
their extraction is more involved for the real emission contributions.
Here, the infrared singularities only become explicit after 
integrating the real radiation matrix elements over the phase space.
To build a general-purpose Monte Carlo program to evaluate 
observables at NLO requires therefore an analytic cancellation of infrared 
singularities before any numerical integration can be performed.

For the task of NLO calculations, 
several systematic and process-independent 
procedures are available 
and have been applied to a variety of processes \cite{Nloprocesses}. 
The two main methods are phase space slicing \cite{gg,ggk} 
and the subtraction methods \cite{Kunszt,cs}.
Except for the phase space slicing technique, all subtraction methods consist 
in introducing terms which are subtracted from the real 
radiation part at each phase space point.
These subtraction terms approximate the matrix element in all singular 
limits and should be sufficiently simple to be integrated over the 
corresponding phase space part analytically.
After this integration, the infrared divergences of the subtraction terms 
become explicit and the integrated subtraction terms can be added to the
virtual corrections, thus yielding an infrared finite result. 
   
One of the widely used subtraction formalisms at NLO 
is the dipole subtraction formalism of Catani and Seymour,
which in its original formulation \cite{cs} deals with 
massless partons in final and/or initial state at NLO.

Another subtraction scheme 
is the antenna formalism \cite{k,antenna,initial}, which constructs 
the subtraction terms from so-called antenna functions.
The antenna functions describe all unresolved partonic (soft and collinear)
radiation between a hard pair of colour-ordered partons, the radiators. 
These functions can be derived systematically 
from physical matrix elements, as shown in \cite{gluino,higgs} 
and can be integrated over the factorised phase space. 
In the antenna subtraction method, the
subtraction terms are constructed from products of 
antenna functions with reduced matrix elements (with fewer partons than the
original matrix element) such that these subtraction terms are acting 
on individual colour-ordered real radiation matrix elements.
So far, this formalism can handle massless partons  
in final and/or initial state at NLO \cite{k,initial} 
and it can treat massless final state parton radiation  
at NNLO \cite{antenna}. 
It is the only formalism where the NNLO formulation has been worked
out in full and applied to evaluate NNLO corrections to jet observables,
namely for $e^+e^- \to 3$ jets \cite{ggg,3jetweinzierl}.

To evaluate hard scattering processes involving {\it massive} 
hard partons in the final state at the next-to-leading 
order level, a subtraction formalism is needed as well in order 
to isolate and cancel infrared divergences among different 
partonic contributions to the cross section. However
the extension from massless partons to massive partons involve 
complications and new features. The kinematics is more involved due to 
the finite value of the parton masses.

Furthermore, QCD radiation from massive partons, although not 
leading to strict collinear divergences, 
(which are regulated by the mass of the massive particle), 
will be proportional 
to $\ln Q^2/M^2$, where $M$ is the parton mass and $Q$ is the typical 
scale of the hard scattering process.
In kinematical configurations where $Q\gg M$, these logarithmically 
enhanced contributions become large and can spoil 
the numerical convergence of the calculation.  
Although these logarithmic terms cancel in the final NLO result 
they appear at intermediate steps of the calculation.
Those can be easily traced though. Indeed, 
these logarithmic enhanced terms are related to a 
process-independent behaviour of the matrix elements; its  
singular behaviour in the massless limit $(M \to 0)$.
This singular behaviour is related to the  {\it quasi-collinear} 
\cite{cdst1} limit of the matrix element.
For massless partons, infrared divergences arise in soft 
and collinear kinematical configurations and in these configurations 
matrix element and phase space obey QCD factorisation formulae 
\cite{fac1}.
Similarly, for massive partons, matrix element and phase space will 
obey factorisation formulae in the soft and quasi-collinear limit. 

The dipole subtraction formalism of Catani and Seymour has been extended 
to include effects of massive partons in \cite{cdst1,cdt2,weinzierl} at NLO.  
An extension of the antenna subtraction method including the 
radiation effects of massive partons has been so far missing. 

It is the purpose of this paper to present an extension of the 
antenna subtraction method to include radiation of final state 
massive fermions in order to be able to evaluate 
the production of massive particles beyond leading order within this formalism.
For the sake of clarity, in this paper we restrict 
ourselves to the kinematical 
situation of a colour-neutral particle decaying into 
one or two coloured massive fermions of equal masses.
In this situation, infrared singularities appear only due to 
final state radiation and only two scales are involved in the problem.
As the radiators are in the final state only, the corresponding 
antenna functions involving those are regarded as final-final antennae.

The basic structure of the extension of the antenna subtraction formalism 
to include massive particles can be carried over to partons 
in the initial state.
However, to deal with the production of massive final states at 
hadron colliders, the construction of the appropriate subtraction 
terms and their integrations  
cannot be solely performed with the ingredients presented in this paper. 
Besides massless initial-initial antenna functions 
derived in \cite{initial}, which involve massless initial state radiators,  
and massive final-final antenna functions 
which will be derived in this paper, antenna functions involving a massless 
initial state and a massive final state are also required.
As in the massless case, those initial-final antenna functions can be obtained 
from the final-final massive antenna functions 
by crossing one massless parton from the final to the initial state. 
Their integration over the corresponding phase space cannot be taken over 
from the final-final case, though.
%As in the massless case, due to kinematical constraints, 
%we expect those to involve less phase space integrals than 
%in the massive final-final case.  

Furthermore, the extension of the antenna subtraction method involving 
massive final state particles is not restricted to Standard Model processes. 
It can also be applied  to the production of supersymmetric particles,
for example. An appropriate treatment of massive final state scalars, 
like squarks could also be realised within this framework.

So far, final-final massless antennae have found some 
interesting applications:  
Starting from the three-parton antenna functions given in \cite{antenna},
a parton shower event generator VINCIA \cite{vincia} has been constructed.
One could therefore envisage a further implementation in VINCIA of the 
massive NLO final-final antennae presented in this paper. A combination 
between NLO calculation in the antenna subtraction formalism 
and parton showers could then become
feasible for processes involving massless and massive final state fermions.
Originally, parton showers based on the antennae approach for final state 
particles had been derived for the event generator ARIADNE 
\cite{ariadne}.
Furthermore, parton showers related to the dipole subtraction 
formalism involving dipole functions   
were considered in
the litterature \cite{Nagy,mcnlo,Weinzierl2} and are already 
implemented \cite{sherpaPS} in SHERPA \cite{sherpa}.

The paper will be organised as follows.
In Section 2, we recall the structure of the antenna subtraction terms at NLO 
for massless final state partons and state how it extends with the
presence of massive partons.
Section 3 presents the formulae for the massive antenna functions while in 
Section 4 a list of all non-vanishing limits for these antennae is given.  
In Section 5 we present the results for the integrated 
massive antenna functions.
To validate our results for the integrated massive quark-antiquark 
antenna function, we recompute the NLO correction to the hadronic decay rate 
in Section 6. Section 7 contains our conclusions.

\section{Antenna subtraction with massive final-final configurations}
\setcounter{equation}{0}
In configurations involving final-final antennae, both radiators are in the
final state. For massless radiators, this case was described in detail at NLO 
in \cite{k} and NNLO in \cite{antenna}.
For massive radiators, the same factorisation formulae 
for the subtraction terms will hold.
Only the ingredients, phase space and antenna functions will be modified to
take into account the mass of radiators.
For clarity reasons, we will keep the same notation as in \cite{antenna}.

The subtraction term 
for an unresolved parton $j$, emitted between massless or massive 
hard final-state radiators $i$ and $k$ reads,

\begin{eqnarray}
{\rm d}\sigma^{S}_{NLO} &=&
{\cal N}\,\sum_{m+1}{\rm d}\Phi_{m+1}(k_{1},\ldots,k_{m+1};q)
\frac{1}{S_{{m+1}}} \nonumber \\ 
 && \sum_{j}\;X^0_{ijk}\,
|{\cal M}_{m}(k_{1},\ldots,{K}_{I},{K}_{K},\ldots,k_{m+1};q)|^2\,
%\nonumber\\ && 
\JET_{m}^{(m)}(k_{1},\ldots,{K}_{I},{K}_{K},\ldots,k_{m+1};q)\,. \nonumber \\ && 
\label{eq:sub1}
\end{eqnarray}
This subtraction term involves the $m$ parton amplitude ${\cal M}_{m}$
 which depends 
only on the redefined on-shell momenta $k_{1},...,k_{I},k_{K},..,k_{m+1}$. 
The momenta $k_{I}$,$k_{K}$ are linear combinations of the momenta 
$k_{i},k_{j},k_{k}$ while the tree-level
antenna function $X^{0}_{ijk}$ depends only on $k_{i},k_{j},k_{k}$, $X$ 
stands for different antenna types denoted (in the massive NLO case) 
by $A,E,D$ or $G$ depending on which radiators are involved.
The jet function $\JET^{(m)}_m$ in eq.(\ref{eq:sub1}) does not depend on the 
individual momenta ${k}_{i}$, $k_j$ and ${k}_{k}$, but only on 
${K}_{I},{K}_{K}$. This function ensures that any 
partonic contribution give rise to an $m$-jet final state. 

The phase space $\d \Phi_{m+1}$ can be factorised as follows,
\begin{eqnarray}
\label{eq:psx3}
\lefteqn{\d \Phi_{m+1}(k_{1},\ldots,k_{m+1};q)  = }
\nonumber \\ &&
\d \Phi_{m}(k_{1},\ldots,{K}_{I},{K}_{K},\ldots,k_{m+1};q)
\cdot 
\d \Phi_{X_{ijk}} (k_i,k_j,k_k;{K}_{I}+{K}_{K})\;.
\end{eqnarray}
For massive radiators, the phase space 
${\rm d}\Phi_{m}$ is the $d$-dimensional ($d=4-2\eps$) 
phase space for $m$ outgoing 
particles with momenta $p_{1},\cdots ,p_{m}$, masses 
$m_{1}, \cdots, m_{m}$ with total
four-momentum $q^{\mu}$. It reads,
\begin{equation}
\label{eq:phi}
\d \Phi_m(k_{1},\ldots,k_{m};q) 
= \frac{\d^{d-1} k_1}{2E_1 (2\pi)^{d-1}}\; \ldots \;
\frac{\d^{d-1} k_m}{2E_m (2\pi)^{d-1}}\; (2\pi)^{d} \;
\delta^d (q - k_1 - \ldots - k_m) \,,
\end{equation}
with
\begin{equation}
E_{i}=\sqrt{|\vec{p}_{i}|^2 +m_{i}^2}.
\end{equation}
$S_{m}$ is a
symmetry factor for identical partons in the final state.
$d\Phi_{X_{ijk}}$ is the NLO antenna phase space, it is proportional 
to the three-particle phase space relevant to a $1\to 3$ decay.
This can be seen by using $m=2$ in the above formula (\ref{eq:psx3}) 
and exploiting the fact that the two-particle phase space is a constant.
The massive NLO antenna phase space  will be derived in Section 5. 

The tree-level antenna function $X^{0}_{ijk}$ 
describes all the situations where parton $j$ is unresolved.
Those are obtained by 
normalising the corresponding colour-ordered three-parton tree-level 
squared matrix elements to the squared matrix element 
for the basic two-parton process, omitting all couplings and colour factors.
Indeed, the normalisation for the antennae is chosen such that 
in the unresolved limits those yield exactly the well-known collinear 
splitting functions and soft eikonal factors.
As such the massless and massive antenna functions 
have mass dimension $-2$ and are defined by 
\cite{antenna},
\begin{eqnarray}
X_{ijk}^0 = S_{ijk,IK}\, \frac{|{\cal M}^0_{ijk}|^2}{|{\cal M}^0_{IK}|^2}\;,
\end{eqnarray}
where $S$ denotes the symmetry factor associated to the antenna, which accounts
both for potential identical particle symmetries and for the presence 
of more than one antenna in the basic two-parton process.
The massive antenna functions will be given below in Section 3
while their unresolved limits will be presented in Section 4.

\section{Massive NLO antenna functions}
\setcounter{equation}{0}
The massive final-final antenna functions 
can be derived from physical matrix elements 
of the same processes as in the massless case 
 \cite{antenna,gluino,higgs}
but considering the final state radiators to be massive. 
More precisely, 
the quark-antiquark antenna function which we denote by $\A$ 
(to be distinguished from $A^{0}_{qg\bar{q}}$)  can be derived 
from $\gamma^{*} \to Q \bar{Q}+g$   with $Q$ being a massive quark 
of mass $m_{Q}$, $g$ is a massless gluon. 
We find three types of massive quark-gluon antennae involving either one or
two massive partons. Those can be derived from the following processes:
($\tilde{\chi} \to \tilde{g} +$ 2 partons). For these final state partons 
being two gluons, we
derive $D^{0}_{Qgg}$ while for these partons being a quark-antiquark pair, 
we derive $\Eqm$ and $\Eqprimem$.
Lastly, we can also define a massive gluon-gluon antenna through the process 
 $H  \to g Q\bar{Q}$, with one of the gluons in the final state emitting a
massive $Q \bar{Q}$ pair. 
All new massive antenna functions are given below with  
$m_{Q}$ denoting the mass of the massive quark $Q$, 
$g$ is a massless gluon and  
the invariant $s_{ij}$ is chosen to be $s_{ij}=2\, p_i\cdot p_j$ making the 
mass dependence in the formulae below explicit. $E_{cm}$ is  
always the rest energy of the decaying particle.
All antennae listed below are known exactly at all orders in $\eps$. 
The $\order{\eps}$ parts have been omitted 
for conciseness only, those are however already needed 
at this next-to-leading order level.
%The $\order{\eps}$ parts are not neglected here.

The quark-antiquark massive antenna function $\A$ reads,
\begin{equation}
\begin{split}
 \A(1,3,2) &= \frac{1}{4 \left(\enen + 2 m_{Q}^2\right)} 4 \left(\frac{2 s_{12}^2}{s_{13} s_{23}}+\frac{2
   s_{12}}{s_{13}}+\frac{2
   s_{12}}{s_{23}}+\frac{s_{23}}{s_{13}}+
   \frac{s_{13}}{s_{23}}\right.\\
& \quad \left.+ m_{Q}^2 \left(\frac{8 s_{12}}{s_{13} s_{23}}-\frac{2
   s_{12}}{s_{13}^2}-\frac{2
   s_{12}}{s_{23}^2}-\frac{2
   s_{23}}{s_{13}^2}-\frac{2}{s_{13}}-\frac{2
   }{s_{23}}-\frac{2 s_{13}}{s_{23}^2}\right)\right.\\
& \quad \left. + m_{Q}^4 \left(-\frac{8}{s_{23}^2}-\frac{8}{s_{13}^2}\right)\right)
  	+\order{\epsilon}.
\end{split}
\end{equation}
This function has been normalised to the two-particle matrix element 
relevant for $\gamma^{*} \to Q\bar{Q}$, whose matrix element squared (omitting 
couplings) is given by 
\begin{equation}
A^{0}_{Q\bar{Q}}(1,2)=4 \left[(1-\eps)\,E_{cm}^2 +2\,m_{Q}^2\right].
\end{equation}

The quark-gluon massive antennae with either two gluons or a
massless quark-antiquark pair in the final state are normalised by the 
two-particle matrix element squared 
relevant for $\tilde{\chi}  \to \tilde{g}g $, with the gluino $\tilde{g}$
 being massive with mass $m_{Q}$, the gluon $g$ being massless. 
This two-particle matrix element squared omitting couplings
reads:
\begin{equation}
X^{0}_{Q g}(1,2)=4\,(1-\eps)\,\left(E_{cm}^2-m_{Q}^2 \right)^2,
\end{equation}
where $X$ can stand for $E$ or $D$ here.
The quark-gluon massive antenna $\D$ with two gluons and a massive 
radiator $Q$ in the final state reads, 
\begin{equation}
\begin{split}
 \D(1,3,4) &= \frac{1}{4 (\enen -m_{Q}^2)^2}
4 \left(
	\left( 9 s_{13} + 9 s_{14} + \frac{4 s_{13}^2}{s_{14}} + \frac{4 s_{14}^2}{s_{13}} + \frac{4 s_{13}^2}{s_{34}} + \frac{2 s_{13}^3}{s_{14} s_{34}} \right. \right. \\
	& \quad \quad \left. \left. + \frac{6 s_{13} s_{14}}{s_{34}} + \frac{4 s_{14}^2}{s_{34}} + \frac{2 s_{14}^3}{s_{13} s_{34}} + 6 s_{34} + \frac{3 s_{13} s_{34}}{s_{14}} + \frac{3 s_{14} s_{34} }{s_{13}} + \frac{s_{34}^2}{s_{13}} + \frac{s_{34}^2}{s_{14}} \right)
	\right. \\
	& \quad \left.
	- m_{Q}^2 \left( 6 + \frac{2 s_{13}^2}{s_{14}^2} + \frac{4 s_{13}}{s_{14}} + \frac{4 s_{14}}{s_{13}} + \frac{2 s_{14}^2}{s_{13}^2} + \frac{6 s_{34}}{s_{13}} + \frac{4 s_{13} s_{34}}{s_{14}^2} \right. \right.\\
	& \quad \quad \left. \left. + \frac{6 s_{34}}{s_{14}} + \frac{4 s_{14} s_{34}}{s_{13}^2} + \frac{2 s_{34}^2}{s_{13}^2} + \frac{2 s_{34}^2}{s_{14}^2} + \frac{2 s_{34}^2}{s_{13} s_{14}} \right)
	\right. \\
	& \quad \left.
	+ 2 \en m_{Q} - \en m_{Q}^3\frac{2  s_{34}}{s_{13} s_{14}} 
+ m_{Q}^4\frac{2 s_{34}}{s_{13} s_{14}} 
\right) + \order{\epsilon}.
 \end{split}
\end{equation}

The quark-gluon massive antenna $\Eqm $ with a pair of massless quarks and a
massive radiator quark $Q$  in the final state reads, 
\begin{equation}
 \Eqm(1,3,4) = \frac{1}{4 (\enen - m_{Q}^2)^2} 4 \left( s_{13} + s_{14} + \frac{s_{13}^2}{s_{34}} + \frac{s_{14}^2}{s_{34}} - 2 \en m_{Q}\right) + \order{\epsilon}.
\end{equation}

In the case where the gluino is massless and the gluon radiates 
a massive quark-antiquark pair, the two-particle matrix element which serves 
to normalise the  quark-gluon massive antenna $\Eqprimem $ 
is different. It reads:
\begin{equation}
E_{q g}(1,2)=4 \en^4\,(1-\eps).
\end{equation}
The quark-gluon massive antenna $\Eqprimem $ 
with a pair of massive quarks and a
massless quark-radiator in the final state reads: 
\begin{equation}
\begin{split}
 \Eqprimem(1,3,4) &= \frac{1}{4 \en^4} \frac{4}{(s_{34} + 2 m_Q^2)^2}\left( s_{13}^2 s_{34} + s_{14}^2 s_{34} + s_{13} s_{34}^2 + s_{14} s_{34}^2 \right. \\
& \quad \left. + m_{Q}^2 \left( 4 s_{13}^2 + 4 s_{13} s_{14} + 4 s_{14}^2 + 6 s_{13} s_{34} + 6 s_{14} s_{34} \right) \right.  \\
& \quad \left. +8 m_{Q}^4 \left(s_{13}+s_{14}  \right) \right) + 
\order{\epsilon}.
\end{split}
\end{equation}
Finally, the gluon-gluon massive antenna where the final state gluon emits a massive quark-antiquark pair reads:
\begin{equation}
 \G(1,3,4) = \frac{1}{4 \en^4} \frac{4}{(s_{34} + 2 m_Q^2)^2}\left( s_{13}^2 s_{34} + s_{14}^2 s_{34} + 4 m_{Q}^2 (s_{13}^2 + s_{14}^2 + s_{13} s_{14} )\right) + \order{\epsilon}.
\end{equation}
This antenna function has been normalised to the two-particle matrix element 
relevant for $H  \to g g$ whose matrix element squared 
is given by,
\begin{equation}
G_{g g}(1,2)=4 \en^4\,(1-\eps).
\end{equation}

\section{Singular limits}
\setcounter{equation}{0}
The antenna functions listed above encapsulate all single unresolved limits
of tree-level QCD matrix elements involving massive final states. 
The factorisation properties of tree-level QCD squared matrix elements 
are well-known  \cite{fac1} for massless and for massive final state
partons.
In the antenna picture, for massless radiators, the unresolved 
and massless parton emitted between those can be either collinear or soft.
In these corresponding unresolved limits, 
defined as in \cite{antenna}  the $(m+1)$-parton matrix element 
squared  factorises into a reduced $m$-parton matrix element 
and the well known soft eikonal factor when a gluon is soft 
and one of the three different Altarelli-Parisi splitting functions \cite{AP} 
when two of the three massless partons are collinear.
When the radiators which emit a massless and unresolved parton between them  
are massive instead, the limiting behaviour of the matrix elements 
has to be changed and the unresolved factors have to be modified accordingly.  
In this section we first recall the single unresolved massless 
factors and present their extension needed 
to deal with massive radiators. 
Finally, we list all non-vanishing limits of the three-parton massive 
antennae defined in Section 3.

\subsection{Single unresolved massless factors}

When a soft gluon ($j$)  
is emitted between two hard and massless partons ($i$ and $k$), the 
 eikonal factor $S_{ijk}$ factorises off the squared matrix element,
\begin{equation}
S_{ijk}\equiv\frac{2s_{ik}}{s_{ij}s_{jk}}.
\label{eq:eikonal}
\end{equation} 
When two massless partons become collinear,
we have different splitting functions  
corresponding to various final state configurations:  
a massless quark splits into a quark and a gluon ($P_{qg\to Q}$), 
a gluon splits into a quark-antiquark pair ($P_{q\bar q\to G}$)  
or a gluon splits into two gluons ($P_{gg\to G}$).
These are given by:
\begin{eqnarray}
P_{qg\to Q}(z)&=&\left(\frac{1+(1-z)^2-\eps z^2}{z}\right) ,\nonumber\\
P_{q\bar{q}\to G}(z)&=&\left(\frac{z^2+(1-z)^2 -\eps}{1-\eps}\right),\nonumber \\
P_{gg\to G}(z)&=& 2\left(\frac{z}{1-z}+ \frac{1-z}{z}+z(1-z)\right)\;.
\label{eq:apkernels}
\end{eqnarray}
In these equations, $z$ is the momentum fraction carried 
by the unresolved parton and the label $q$ present in these splitting 
functions 
can stand for a quark or an antiquark:  
$P_{qg\to Q}= P_{\bar{q}g\to \bar{Q}}$ by charge conjugation.
$Q$ or $G$ appearing in the collinear splitting functions denotes the 
parent particle of the two collinear partons $i$ and $j$.

\subsection{Single unresolved massive factors}
When massive partons are present in the final state, those  
can be soft but they cannot be strictly collinear, the mass is regularizing 
the collinear divergence.
The relation between the massive matrix element squared and the splitting
functions needs to be extended from massless to massive. Similar 
factorisation formulae as in the massless case will hold provided the
collinear limit is generalized~\cite{cdst1}
 to the {\it quasi-collinear} limit.

The limit when a massive parton $P$ of momentum $p^{\mu}$ decays 
quasi-collinearly into two massive partons $j$ and $k$ is defined by,
\begin{equation}
p_{j}^{\mu} \to z \, p^{\mu}, \,p_{k}^{\mu} \to (1-z) \, p^{\mu},
\end{equation}
\begin{equation}
p^2=m_{(jk)}^{2}.
\end{equation}
with the constraints,
\begin{equation}
p_{j}\cdot p_{k},m_{j},m_{k},m_{jk} \to 0
\end{equation}
at fixed ratios,
\begin{equation}
\frac{m_{j}^2}{p_{j}\cdot p_{k}},
\frac{m_{k}^2}{p_{j}\cdot p_{k}},
\frac{m_{jk}^2}{p_{j}\cdot p_{k}}.
\end{equation}

The key difference between the massless collinear limit and the quasi-collinear
limit is given by the constraint that the on-shell masses squared have to be
kept of the same order as the invariant mass $(p_{j}+p_{k})^2$, 
with the latter becoming small.
In this corresponding quasi-collinear limits, 
the $(m+1)$-parton matrix element 
squared factorises into a reduced $m$-parton matrix element 
and unresolved massive factors.
 
More precisely, the single unresolved massive factors 
in which the real matrix elements factorise 
in the soft and quasi-collinear limits 
are generalizations of the massless soft and collinear Altarelli-Parisi 
splitting functions defined above.
The massive splitting functions denoted by 
${P}_{ij \to (ij)}(z,\mu_{ij}^2)$ for parton $(ij)$ splitting into partons 
$i$ and $j$ in $d$-dimensions ($d=4-2\eps$) will be given below.
In these expressions, $z$ is the momentum fraction carried by 
the unresolved parton becoming quasi-collinear 
to the massive parton. All the mass dependence can
be parameterized 
by $\mu_{jk}=(m^2_j +m^2_k)/[(p_j +p_k)^2 -m^2_{(jk)}]$.
The 
number of gluon polarizations is chosen within the dimensional regularisation
procedure as $d-2$ . 
Those generalized massive splitting functions were also given 
in the appendix of \cite{cdt2}
and read,
\begin{equation}
 \begin{split}
 {\splitqgq (z,\mu^{2}_{qg})} 
&= \left( \frac{1+(1-z)^2-\eps z^2}{z} 
- 2 {\mu_{Qg}^2} \right)\\
& \quad \text{where $\mu^{2}_{qg}$ is defined by 
$ \mu^2_{qg}=\frac{m^2_Q}{s_{qg}}$ },\\
  {\splitqqbarg (z,\mu^2_{q\bar{q}})} &= \left(\frac{z^2 +(1-z)^2 -\eps
-\mu^2_{q\bar{q}}}{1-\epsilon}\right)
\nonumber \\
& \quad \text{where $\mu^2_{q\bar{q}}$ is defined by 
$ \mu^2_{q\bar{q}}=\frac{2 m^2_Q}{[s_{q\bar{q}}+2\,m^2_{Q}]}$ }.\\
\end{split}
\end{equation}
The gluon-gluon splitting function ${\splitggg (z)}$ is left unchanged.
The generalized soft eikonal factor ${S}_{ijk}(m_{i},m_{k})$ 
depends on the invariants $s_{lm}=2\, p_l\cdot p_m$ built with 
the partons $i,j$ and $k$ 
but also on the masses of partons $i$ and $k$, 
$m_{i}$ and $m_{k}$ respectively.
It is given by,
\begin{equation}
{\softeikonal_{ijk}}(m_{i},m_{k})= \frac{2 s_{ik}}{s_{ij}s_{jk}} 
- \frac{2 m_i^2}{s_{ij}^2} - \frac{2 m_k^2}{s_{jk}^2}.
\end{equation}

\subsection{Singular limits of the massive NLO antenna functions} 
We list here the non-vanishing soft, collinear and quasi-collinear limits 
of the massive antenna functions given in Section 3.
The singular limits of \A are
\begin{equation}
 \begin{split}
 \A \left(1,3,2 \right) & \xrightarrow{3_g \rightarrow 0} 
\softeikonal_{132}(m_Q,m_Q),\\
\A \left(1,3,2 \right) & \xrightarrow{3_g \parallel 1_Q} \frac{1}{s_{13}}
\splitqgq(z,\mu^2_{qg}),\\
\A \left(1,3,2 \right) & \xrightarrow{3_g \parallel 2_{\bar{Q}}} \frac{1}{s_{23}}
\splitqgq(z,\mu^2_{qg}).\\
\end{split}
\end{equation}

%*Due to the symmetry of $\A$ under exchange of quark and antiquark 
%*the limit for $3_g \parallel 2_{\bar{Q}}$ is analogous.\\
For \D we have
\begin{equation}
\begin{split}
\D \left(1,3,4 \right) & \xrightarrow{3_g \rightarrow 0} 
\softeikonal_{134}(m_{Q},0),\\
\D \left(1,3,4 \right) & \xrightarrow{3_g \parallel 1_Q} \frac{1}{s_{13}}
\splitqgq(z,\mu^2_{qg}),\\
\D \left(1,3,4 \right) & \xrightarrow{4_g \parallel 1_Q} \frac{1}{s_{14}}
\splitqgq(z,\mu^2_{qg}),\\
\D \left(1,3,4 \right) & \xrightarrow{3_g \parallel 4_g} \frac{1}{s_{34}} 
\splitggg(z).\\
\end{split}
\end{equation}

The only non-vanishing singular limit of \Eqm is the collinear massless limit 
of the massless quark-antiquark pair,
\begin{equation}
\Eqm \left(1,3,4 \right) \xrightarrow{3_{q'} \parallel 4_{\bar{q}'}} 
\frac{1}{s_{34}}  \splitqqbarg(z),
\end{equation}
while the only non-vanishing limit of $\Eqprimem$ is, 
\begin{equation}
\Eqprimem \left(1,3,4 \right) \xrightarrow{3_{Q'} \parallel 4_{\bar{Q}'}} 
\frac{1}{s_{34}+2 m_{Q}^2} \splitqqbarg(z,\mu^2_{q\bar{q}}).
\end{equation}
Finally \G does also have a quasi-collinear quark-antiquark limit given by
\begin{equation}
 \G \left(1,3,4 \right) \xrightarrow{3_{Q'} \parallel 4_{\bar{Q}'}} 
\frac{1}{s_{34}+2 m_{Q}^2} \splitqqbarg(z,\mu^2_{q\bar{q}}).
\end{equation}

\section{Integrated massive NLO antenna functions}
\setcounter{equation}{0}
In this section we present the results for 
the integration over the appropriate antenna phase space
of the massive final-final three-parton antenna functions 
defined in Section 3.

As in the massless case, the phase space associated to the 
subtraction term given in eq.(\ref{eq:psx3}) can be factorised.
In a first step, we present the phase space factorisation 
of a three-particle phase space with three massive particles into a
two-particle phase space and an antenna phase space. The latter is proportional 
to a three particle massive phase space. We need to consider two cases:
three particle final state with one massless parton and 
either two massive partons  of equal mass, or one massive parton and two 
massless partons.

\subsection{Phase space factorisation}

In the rest frame of a decaying photon of momentum  $q$, 
the $d$-dimensional two-particle phase space ${\rm d}\Phi_{2}(p_{I},p_{K};q)$ 
given in eq.(\ref{eq:phi}) for two outgoing particles 
$I$ and $K$ with momenta $p_{I}$ and $p_{K}$ and masses $m_{I}$ and $m_{K}$ 
reads:
\begin{equation}
 d \Phi_2((p_{I},p_{K};q)=
(2 \pi)^{2-d} \left(\frac{1}{2}\right)^{d-1} (\en ^2)^{1-d/2} 
\left( (\en^2-m_I^2-m_K^2)^2 - 4 m_I^2 m_K^2\right)^{\frac{d-3}{2}}
{\rm d}\Omega_{d-1},
\end{equation}
with
\begin{equation}
\int{\rm d}\Omega_{d}=\frac{2 \pi^{d/2}}{\Gamma(d/2)},
\end{equation}
where we have used
\begin{equation}
\int \frac{{\rm d}^{d-1}p_{I}}{2 E_{I}}=\frac{1}{2}\int{\rm d}E_{I} 
|\vec{p}_{I}|^{d-3}{\rm d}\Omega_{d-1}.
\end{equation}
In this last equation, 
$\Omega_{d-1}$ parameterizes the solid angle of the $d-1$ 
components of $\vec{p}_{I}$ in $d-1$ dimensions.
The three particle phase space ${\rm d}\Phi_{3}(p_{i},p_{j},p_{k};q)$ 
%given in eq.(\ref{eq:phix3}) 
may be written as the product 
of a two particle phase space ${\rm d}\Phi_{2}(p_{I},p_{K};q)$ defined above 
and an antenna phase space $d\Phi_{X_{ijk}}$ 
which depends on the momenta $p_{i},p_{j}$ and $p_{k}$ 
and on the masses $m_{i},m_{j}$ and $m_{k}$.

Using spherical coordinates in $d$ dimensions, we have
\begin{equation}
\int \frac{{\rm d}^{d-1}p_{i}}{2 E_{i}}\;
 \frac{{\rm d}^{d-1}p_{j}}{2 E_{j}}=
\int \frac{1}{4}\left[ |\vec{p}_{i}||\vec{p}_{j}|\sin{\theta}\right]^{d-3}
{\rm d}E_{i}\,{\rm d}E_{j}\,{\rm d \theta}\,
{\rm d}\Omega_{d-2}{\rm d}\Omega_{d-1}
\end{equation}
with $\theta$ being the angle between the vectors $\vec{p}_{i}$ and 
$\vec{p}_{j}$. $\Omega_{d-1}$, $\Omega_{d-2}$ parameterize the solid angles 
of the $d-1$ and $d-2$ components of the vectors 
$\vec{p}_{i}$ and $\vec{p}_{j}$ respectively, in $d-1$ dimensions.
We find,
\begin{equation}
\label{eq:Phi3}
\int{\rm d}\Phi_{3}(p_{i},p_{j},p_{k};q)=\int {\rm d}\Phi_{2}(p_{I},p_{K};q)
\times d\Phi_{X_{ijk}}
\end{equation} 
with
the antenna phase space $d\Phi_{X_{ijk}}$ given by,
\begin{equation}
\label{eq:phix}
\begin{split}
\int d \Phi_{X_{ijk}} & (s_{ij},s_{jk},s_{ik}) =\\
	&\,(2 \pi)^{1-d} \frac{2 \pi^{d/2-1}}
{\Gamma\left(\frac{d}{2}-1\right)}
 \frac{1}{4} \left( (\en^2-m_I^2-m_K^2)^2 - 4 m_I^2 m_K^2\right)^{\frac{3-d}{2}}
\\
	&\int d s_{ij}\, ds_{jk}\,ds_{ik}\, \delta( \en^2-m_i^2-m_j^2-
m_k^2-s_{ij}-s_{jk}-s_{ik})\\
	&[4\,\Delta_3(p_{i},p_{j},p_{k})]^\frac{d-4}{2} \;
\theta(\Delta_3(p_{i},p_{j},p_{k})). 
\end{split}
\end{equation}
The masses $m_{I}$ and $m_{K}$ appearing in this equation are
combinations of the masses $m_{i},m_{j}$ and $m_{k}$ and  the integral over   
$\Omega_{d-2}$ has been performed.
The function $\Delta_3(p_{i},p_{j},p_{k})$  
is the Gram determinant for massive particles of momenta 
$p_{i},p_{j},p_{k}$ given in terms of invariants $s_{ij}=2p_i \cdot p_j$ 
and masses $m_{i},m_{j},m_{k}$ by,
\begin{equation}
\Delta_3(p_{i},p_{j},p_{k}) = 
\frac{1}{4}\left(s_{ij}s_{ik}s_{jk} - m_i^2 s_{jk}^2-m_k^2s_{ij}^2
-m_j^2 s_{ik}^2 + 4 m_i^2 m_j^2 m_k^2 \right)
\end{equation}

As in the massless case, the factorisation of the phase space 
is exact. The expression for the antenna phase space 
$d\Phi_{X_{ijk}}$ given in eq. (\ref{eq:phix}) can also be found  
in \cite{cdst1,weinzierl}, where it was called a dipole phase space.  
At NLO, both massless and massive dipole and antenna phase spaces are the same.
The NLO massive antenna functions differ however from the NLO massive dipoles
functions given \cite{cdst1} as did the NLO massless antenna \cite{antenna} 
from the massless dipole functions \cite{cs}.  

In the following, we will only consider two kinematical configurations of the final state particles. 
For the case where one of mass vanishes ($m_{j}=0$) 
and the two other masses are equal ($m_i=m_k=m_{Q}$), i.e  
$m_{I}=m_{K}=m_{i}$ in eq.(\ref{eq:phix}), 
we use the following parameterization,
\begin{equation}
\label{eq:Phix1}
\begin{split}
 d \Phi_{X_{ijk}}^{(m,0,m)} &= \frac{(4 \pi)^{\epsilon-2}}{\Gamma(1-\epsilon)} \left( \enen \right)^{1-\epsilon} \left( r_0 \right)^{2-2\epsilon} 2^{2 \epsilon -1}\\
& \int_0^1 dr r^{1-2\epsilon}(1-r)^{-\epsilon+\frac{1}{2}} \left( 1- r_0 r\right)^{-\frac{1}{2}} \int_{-1}^1 ds \left(1-s^2\right)^{-\epsilon},\\
& r_0 = 1 - \frac{4 m_{Q}^2}{\enen},\\
& s_{ij} = \frac{1}{2} \enen r_0 r \left( 1-s \sqrt{\frac{r_0(1-r)}{1-r_0 r}} \right),\\
& s_{jk} = \frac{1}{2} \enen r_0 r \left( 1+s \sqrt{\frac{r_0(1-r)}{1-r_0 r}} \right),
\end{split}
\end{equation}
and for the case where two particles are massless $m_i=m_{j}=0$ and the third
one is massive, $m_{k}=m_{Q}$, in which case $m_{I}=m_{i}=0$, $m_{K}=m_{k}$ 
in eq.(\ref{eq:phix}), we use,
\begin{equation}
\label{eq:Phix2}
 \begin{split}
 d \Phi_{X_{ijk}}^{(0,0,m)} &= \frac{(4 \pi)^{\epsilon-2}}{\Gamma(1-\epsilon)} \left( \enen \right)^{1-\epsilon} \left(u_0\right)^{2 - 2 \eps}\\
& \quad \int_0^1 du u^{1-2\eps} (1-u)^{1-2\eps}(1-u_0 u)^{-1+\eps} \int_0^1 dv v^{-\eps} (1-v)^{-\eps},\\
& \quad u_0 = 1 - \frac{m_{Q}^2}{\enen},\\
& \quad s_{ij} = \enen u_0^2 \frac{u(1-u)v}{(1-u_0 u)},\\
& \quad s_{ik} = \enen u_0 (1-u).\\
%& \quad s_{jk} = \enen u_0\,u \frac{1-u_0(1-u)v}{(1-u_0 u)}.\\
\end{split}
\end{equation}

\subsection{Integrals}

Following the extension of the integration by parts method \cite{IBP1,IBP2} in \cite{higgsnnlo,4particle}, 
to reduce the number of real phase space integrals,
we have expressed all invariants in terms of massive propagators  and 
expressed the three on-shell conditions $p_i^2 = m_i^2$, $i=1,2,3$ 
as cut propagators. The reduction to master integrals using 
the Laporta algorithm \cite{Laporta} was done independently, 
once with an in-house implementation in \form  \enspace \cite{form} 
and once with the mathematica package 
\fire \enspace \cite{firepaper}. We find five master integrals, four of which can be evaluated 
in terms of hypergeometric functions for arbitrary $\epsilon$.
For the last one, only the expansion
up to order $\eps^2$ will be given. Differential equation techniques 
\cite{remiddiDE,gehrmann-remiddi} are used to compute this master integral. 

The integrated antennae can be separated in two categories corresponding 
to the phase space parameterizations used. 
The integrated antennae with two massive particles 
are obtained by integrating the antennae with two massive final state 
partons presented in Section 3 over the antenna phase space given 
in eq.(\ref{eq:Phix1}).
Those can be expressed in terms of two master integrals $I_{i}^{(m,0,m)}$ 
(i=1,2) and are given exactly at any order in $\eps$ by,
\begin{equation}
 \begin{split}
  \Aint &= \frac{1}{\enen \eps (1-2\eps) r_0 (1-r_0) (3-r_0-2\eps)}\\
&\left( \left(2 (1-\eps)(-15+8 r_0-r_0^2 +\eps(28-12 r_0 + 4 r_0^2) 
- \eps^2 (12+4 r_0 + 4 r_0^2) + 8 \eps^3 r_0)\right) \intphimmzero 
	\right. \\ 
& + \left. \frac{1}{\enen} 24 (1-\eps)^2 (3-r_0-\eps(2+r_0)+2 \eps^2 r_0) \intsmmzero \right),\\
\Eqprimemint &= \frac{1}{\enen} \left( \frac{2 r_0 (3+r_0 -4 \eps r_0)}
{3(1-r_0)} \intphimmzero  
- \frac{1}{\enen} 4 \frac{(1 + r_0 - 2 \eps r_0)}{(1-r_{0})} \intsmmzero \right) ,\\
\Gint &= \frac{1}{\enen} \left( \frac{8 (1-\eps) r_0^2}{3 (1-r_0)}
\intphimmzero -\frac{1}{\enen} 8 (1-\eps)\frac{r_0}{(1-r_{0})} \intsmmzero \right). \\
\end{split}
\label{eq:Aint}
\end{equation}
The two master integrals $I_{i}^{(m,0,m)}$ are given exactly below.
$\intphimmzero$ is the integrated phase space measure and is given by 
\begin{eqnarray}
\intphimmzero &=& \int d \Phi_{X_{ijk}}^{(m,0,m)} \nonumber \\
&=&(\enen)^{1- \eps} r_0^{2-2\eps} 2^{-2\eps} \pi^{-2+\eps} 
\frac{\Gamma(2-2\eps) \Gamma(3-3\eps)}{\Gamma(6-6\eps) }
\gaussf{\frac{1}{2}}{2-2\eps}{\frac{7}{2}-3\eps}{r_0}\,, \nonumber \\
\end{eqnarray}
and 
\begin{eqnarray}
\intsmmzero &=& \int d \Phi_{X_{ijk}}^{(m,0,m)} (s_{ij}) \nonumber \\
&=& (\enen)^{2- \eps} r_0^{3-2\eps} 2^{1-2\eps} \pi^{-2+\eps} 
\frac{\Gamma(3-2\eps) \Gamma(4-3\eps)}{\Gamma(8-6\eps)}
\gaussf{\frac{1}{2}}{3-2\eps}{\frac{9}{2}-3\eps}{r_0}\,, \nonumber \\
\end{eqnarray}
with $p_{i}^2=m_{i}^2$ and $p_{j}^2=0$.

%A crucial check on the correctness of $\Aint$ will be presented in Section 6.

The integrated antennae with one massive particle are obtained 
by integrating the antenna functions with one massive final state 
presented in Section 3 with the phase space
given in eq.(\ref{eq:Phix2}) (with $\mu = \frac{m_{Q}}{\en}$).
Those can be expressed in term of three master integrals $I_{i}^{0,0,m}$ 
(i=1,..3) as follows,
\begin{equation}
 \begin{split}
\Dint &= \frac{1}{\enen} \frac{1}{\eps^2}\frac{1}{(1-\eps)(1-2\eps)\mu^2 (1-\mu)^2(1+\mu)^3} \left( 12 \mu^4 (1+\mu)\right.\\
&\quad \left. + 2 \eps (1+\mu)(1-6\mu^2-28\mu^4) - \eps^2 (7 + 7 \mu -42\mu^2-40\mu^3-111\mu^4-125\mu^5)\right.\\
& \quad \left. +2 \eps^3 (5+3\mu-31\mu^2-22\mu^3-44\mu^4-59\mu^5) \right.\\
& \quad \left. -\eps^4 (1+\mu)(7-8\mu-38\mu^2+30\mu^3-19\mu^4)\right.\\
& \quad \left.+2\eps^5(1+\mu)^2(1-3\mu-\mu^2+9\mu^3) \right) \intphizerozerom\\
& \quad + \frac{1}{\enenen}\frac{1}{\eps^2}\frac{1}{(1-2\eps)\mu^2(1-\mu)^2(1+\mu)^3}\left(12 \mu^2 (1+\mu)-2\eps(1+\mu)(3+22\mu^2)\right.\\
& \quad \left. + \eps^2 (15+15\mu+65\mu^2+77\mu^3)-\eps^3(1+\mu)(15-12\mu+31\mu^2)\right.\\
& \quad \left. + 2\eps^4(1+\mu)(3-6\mu-7\mu^2) \right) \intszerozerom\\
& \quad - \enen \frac{1}{1-\eps} \frac{2 \mu}{1+\mu} \left( \mu (1+2 \mu ) +\eps (1-\mu^2)-\eps^2(1+\mu)^2\right) \intdiff,\\
\Eqmint &= \frac{2}{\eps \enen (1-\mu^2)^2} \left(\left(-2 \mu^2 +\eps(3
\mu^2 - \mu)\right) \intphizerozerom - \frac{2 - 3 \eps}{\enen} 
\intszerozerom \right).\\
\end{split}
\end{equation}

The master integral $I_{1}^{(0,0,m)}$ corresponding to the 
integrated phase space  and the master integral $I_{2}^{(0,0,m)}$ 
are given exactly by,
\begin{eqnarray}
\intphizerozerom &=& \int d \Phi_{X_{ijk}}^{(0,0,m)}  \nonumber \\
&=& (\enen)^{1-\eps} u_0^{2-2\eps} 2^{-4+2\eps} \pi^{-2+\eps}
\frac{\Gamma(2-2\eps)
  \Gamma(1-\eps)}{\Gamma(4-4\eps)}\gaussf{1-\eps}{2-2\eps}{4-4\eps}{u_0}\,,
\nonumber \\
\end{eqnarray}
\begin{eqnarray}
\intszerozerom &=& \int d \Phi_{X_{ijk}}^{(0,0,m)} (s_{ik}) \nonumber \\
&=& (\enen)^{2-\eps} u_0^{3-2\eps} 2^{-5+2\eps} \pi^{-2+\eps} 
\frac{\Gamma(2-2\eps) \Gamma(1-\eps)}{\Gamma(4-4\eps)}
 \gaussf{1-\eps}{2-2\eps}{5-4\eps}{u_0}\,, \nonumber \\
\end{eqnarray}
with $p_{i}^2=0,p_{k}^2=m_k^2$.

The third master integral $I_{3}^{(0,0,m)}$ is defined as follows,
\begin{eqnarray}
\intdiff &=& \int \frac{1}{s_{ik} s_{jk}} d \Phi_{X_{ijk}}^{(0,0,m)} 
\nonumber \\
	&=&\frac{(\enen)^{-1-\eps} (4 \pi)^{-2+\eps} u_0^{-2\eps}}
{\Gamma(1-\eps)} \int_0^1 \int_0^1 du dv \frac{u^{-2\eps} (1-u)^{-2\eps} 
(1-u_0 u)^{\eps} v^{-\eps} (1-v)^{-\eps}}{1-u_0 u - (1-u) u_0 v}.\nonumber \\
\end{eqnarray}

To get the integrated quark-gluon antenna denoted by \Dint up to finite terms, 
only the constant term of the integral $I_{3}^{(0,0,m)}$ is needed.
The evaluation of this integral 
yields hypergeometric functions of argument $u_0$. 
The expansion around $\eps=0$ of these hypergeometric functions 
leads to one-dimensional harmonic polylogarithms \cite{huber1,huber2,maitre},
$HPL$-functions, denoted here by $H$. 
The constant term in $I_{3}^{(0,0,m)}$ yields,  
\begin{equation}
 \left. \intdiff \right|_{\eps=0} = \frac{1}{\enen}\frac{1}{16 \pi^2 u_0}\left( \hpl{1,1}{u_0} + \hpl{2}{u_0}\right).
\end{equation}
 
Foreseeing that the results we obtained for the integrated NLO massive 
antenna functions could be used as an input for 
a further development of NNLO antenna subtraction involving 
massive radiators we derived the order $\eps$ and $\eps^2$ of 
this master integral $I_{3}^{(0,0,m)}$.  

For this purpose,
we derived differential equations for $I_{3}$ with respect to $E_{cm}^2$ and 
$u_{0}$.  The equations obtained are homogeneous with respect 
to $E_{cm}^2$ and inhomogeneous with respect to $u_{0}$.  
In the differential equation for $I_{3}$ with respect to $u_{0}$ 
the master integrals $I_{1}$ and $I_{2}$ 
appear as coefficients of the inhomogeneous part of the equation.
By inserting their known results, we could solve the differential 
equation for $I_{3}$ order by order in $\eps$.
Some care has to be taken however.
These master integrals $I_{1}, I_{2}$ and $I_{3}$ are obtained after 
an integration over the antenna phase space $d \Phi_{X_{ijk}}^{(0,0,m)}$.
The differential equation with respect to $u_{0}$ however 
concerns integrals over the full three-parton phase space. 

As we saw in eq. (\ref{eq:Phi3}) the latter phase space 
can be written as the product of the antenna phase space 
and a two-particle phase space.
However, since the two particle phase space depends on $u_0$, special
care has to be taken solving the differential equation for $u_{0}$.
Solving this differential equation will yield the wanted master integral 
$I_{3}$ multiplied by the two-particle phase space 
with one massive and one massless particle denoted by $\Phi_{2}(m,0)$. 
The master integrals $I_{1}$ and $I_{2}$ defined above will not appear as such 
in the differential equation but those appear as multiplied 
by the two-particle phase space $\Phi_{2}(m,0)$.    

More precisely, the two-particle phase space 
with one massive and one massless particle reads,
\begin{equation}
\Phi_{2}(m,0)=2^{-3+2\eps}(\pi)^{-1+\eps}\frac{\Gamma(1- \eps)}
{\Gamma(2-2\eps)}(u_{0})^{1-2 \eps}\,(E_{cm}^2)^{-\eps}.
\end{equation}
The master integrals $I_{1}(u_{0})$ and $I_{2}(u_{0})$ 
appearing in the differential equation for $I_{3}$ are obtained 
as the product of the original master integrals $I_{1}$ and $I_{2}$ 
with the two particle phase space measure $\Phi_{2}(m,0)$. 
Those take the form, 
\begin{eqnarray}
I_{1}(u_{0}) &=& O_{cc}\,(E_{cm}^2)^{1-2 \eps}\,(u_{0})^{3-4 \eps}\,
\gaussf{1-\eps}{2-2\eps}{4-4\eps}{u_0} \nonumber \\ 
I_{2}(u_{0}) &=& O_{cc}\,(E_{cm}^2)^{2-2 \eps}\,(u_{0})^{4-4 \eps}\,
\gaussf{1-\eps}{2-2\eps}{5-4\eps}{u_0}\nonumber \\ 
\end{eqnarray}
with the overall normalisation factor $O_{cc}$  given by,
\begin{equation}
O_{cc}=2^{-7 +4\eps}\,(\pi)^{-3+2\eps}
\frac{\Gamma(1- \eps)^2}{\Gamma(4-4 \eps)}.
\end{equation}
With this notation, the master integral $I_{3}$ is given up to order $\eps^2$ 
by,
\begin{equation}
I_{3}=\frac{1}{\Phi_{2}(m,0)} \, O_{cc}\,\hat{I}(E_{cm}^2)^{-1-2 \eps}
\end{equation}
with,
\begin{eqnarray}
 \hat{I} &=& \left( 6 \hpl{2}{u_0} + 6 \hpl{1,1}{u_0} - 2 \eps \left( 16 \hpl{2}{u_0} + 6 \hpl{3}{u_0} + 16 \hpl{1,1}{u_0}\right. \right.\nonumber \\ &&
\left. \left.   - 3 \hpl{1,2}{u_0} + 12 \hpl{2,0}{u_0} + 3 \hpl{2,1}{u_0} + 12 \hpl{1,1,0}{u_0} - 6 \hpl{1,1,1}{u_0} \right) \right. \nonumber \\ &&
\left. +2 \eps^2 \left( 16 \hpl{2}{u_0} + 32 \hpl{3}{u_0} + 12 \hpl{4}{u_0} + 16 \hpl{1,1}{u_0} - 16 \hpl{1,2}{u_0}  \right. \right. \nonumber \\ &&
\left. \left.  - 6 \hpl{1,3}{u_0} + 64 \hpl{2,0}{u_0} + 16 \hpl{2,1}{u_0} + 24 \hpl{3,0}{u_0} + 6 \hpl{3,1}{u_0}\right. \right. \nonumber \\ &&
\left. \left.  + 64 \hpl{1,1,0}{u_0} - 32 \hpl{1,1,1}{u_0} + 9 \hpl{1,1,2}{u_0} - 12 \hpl{1,2,0}{u_0}  \right. \right. \nonumber \\ &&
\left. \left. -3\hpl{1,2,1}{u_0}+48\hpl{2,0,0}{u_0} +12\hpl{2,1,0}{u_0}-9\hpl{2,1,1}{u_0}\right. \right. \nonumber \\ &&
\left. \left.+ 48 \hpl{1,1,0,0}{u_0}-24\hpl{1,1,1,0}{u_0}+9\hpl{1,1,1,1}{u_0} \right) \right).
\end{eqnarray}

\section{Check of \Aint}
\setcounter{equation}{0}
Our new result for the integrated NLO massive quark-antiquark antenna 
$\Aint$ given in eq.(\ref{eq:Aint}) can be tested with expressions known 
in the literature. 
The integrated antenna function $\Aint$ can be regarded as  
the order ${\as}$ part of the real radiation 
correction to the decay rate of a virtual photon into a massive 
quark-antiquark pair, $\gstar \rightarrow Q \bar{Q}$.
By adding the real corrections obtained with $\Aint$ 
to the order ${\as}$ part of the virtual corrections for this decay rate, known in the literature, \cite{bernreuther,Jersak,Schwinger} 
we are able to reproduce the known result 
for the total hadronic decay width at order ${\as}$ 
\cite{Schwinger,Kuhn}.
The details of the comparison are given below.
All formulas are restricted to vector coupling only 
and one heavy flavour $f$ of quarks $Q$. 

The decay width of a virtual photon ($\gamma ^{*}$) 
into a quark-antiquark pair is given by,
\begin{equation}
\Gamma^{had}_{\gstar} = \frac{1}{2}  \alpha \en N_C \,Q_f^2 \,r^V_{NS}(f).
\label{eq:Gammahad}
\end{equation}
with the colour factor $C_{F}$ given by $C_{F}=\frac{N_{c}^2-1}{2 N_{c}}$, 
$N_{c}$ the number of colours
and,
\begin{eqnarray}
r^V_{NS}(f) &=& v \frac{3-v^2}{2} \left[ 1+ \cf \frac{\as}{\pi} K_V \right]\\
K_V &=& \frac{1}{v}\left( A(v)+ \frac{\frac{33}{24}+\frac{22}{24}v^2-\frac{7}{24}v^4}{1-v^2/3} \ln \frac{1+v}{1-v} + \frac{\frac{5}{4}v-\frac{3}{4}v^3}{1-v^2/3}\right)\\
A(v) &=& (1+v^2) \left[ \li\left(\left[\frac{1-v}{1+v}\right]^2\right) + 2 \li
  \left(\frac{1-v}{1+v}\right) + \ln \frac{1+v}{1-v} \ln \frac{(1+v)^3)}{8v^2}
\right]\nonumber \\ &&
+ 3 v \ln \frac{1-v^2}{4 v}-v \ln v.
\end{eqnarray}
In this equation, $r_0=1-\frac{4m^2}{\enen}$ is used 
as the dimensionless variable and $v=\sqrt{r_0}$.
This result   
is obtained from the decay width of $Z$ boson 
into a quark-antiquark pair \cite{Kuhn} by appropriate coupling replacements.
It can be reproduced by computing 
the vertex correction $\gammavertex$ and the real radiation correction 
$\gammareal$ separately. The corresponding formulae will be given below.

The vertex correction $\gammavertex$ is given in \cite{bernreuther,Schwinger} 
in the form of the QCD vertex amplitude $V^{\mu}_{c_1 c_2}$ 
corresponding to the decay of a virtual photon with momentum $p_1 + p_2$
 into a quark and an antiquark with colours $c_1$ and $c_2$
 (and $\sigma^{\mu \nu} = \frac{i}{2} \left[\gamma^{\mu},\gamma^{\nu}\right]$)
\begin{eqnarray}
V^{\mu}_{c_1 c_2}(p_1,p_2) &=& \bar{u}_{c_1}(p_1) \Gamma^{\mu}_{c_1 c_2}(p_1,p_2)v_{c_2}(p_2),\\
\Gamma^{\mu}_{c_1 c_2}(p_1,p_2)&=&-i \sqrt{4 \pi \alpha} Q_f \delta_{c_1 c_2}
\left( F_1\left(\frac{\enen}{m^2}\right) \gamma^{\mu} + \frac{1}{2 m}
  F_2\left(\frac{\enen}{m^2}\right) i \sigma^{\mu \nu} (p_1
  +p_2)_{\nu}\right)\,,\nonumber \\ 
\end{eqnarray}
where $Q_{f}$ is the quark charge in units 
of charge $e=\sqrt{4 \pi \alpha}$ for one flavour $f$.
The hadronic vertex correction $\gammavertex$ then reads,
\begin{equation}
\gammavertex = \frac{1}{2 \en} \frac{1}{2} \sum_{spins, colours} V^{\mu}_{c_1 c_2} V_{\mu c_3 c_4} \Phi_2\\.
\end{equation}
$\Phi_2$ is the integrated two-particle phase space for two massive particles given by,
\begin{equation}
\Phi_2 (m,m)= 2^{-3+2\eps}(\pi)^{-1+\eps}
\frac{\Gamma(1- \eps)}{\Gamma(2-2\eps)} (r_{0})^{1/2 - \eps} (\enen)^{-\eps}
\end{equation}
and the amplitude squared is,
\begin{eqnarray}
\sum_{spins, colours} V^{\mu}_{c_1 c_2} V_{\mu c_3 c_4} =
 N_C \,(4 \pi \alpha Q_f^2)\, \enen \left(\abs{F_1}^2
 ( 2 (3 - r_0 - 2 \epsilon) \right.  \nonumber \\
 \quad \left. + \abs{F_2}^2 (\frac{2}{1-r_0} (3-2r_0 - 2 \epsilon (1-r0))+\Re\left(F_1 F_2^*\right) (4 (3 - 2 \epsilon))\right).
\end{eqnarray}
The form factors $F_1$ and $F_2$ are given up to order $\as$ and $\epsilon^0$ 
by,
\begin{eqnarray}
F_1 &=& 1 + C(\epsilon) \frac{\as}{2 \pi} \mathcal{F}^{1l}_{1,R}\\
F_2 &=& C(\epsilon) \frac{\as}{2 \pi}\mathcal{F}^{1l}_{2,R}
\end{eqnarray}
with the normalisation factor $C(\epsilon)$ given by,
\begin{equation}
C(\epsilon) = (4 \pi)^{\epsilon} \Gamma(1+\epsilon) \left( \frac{\mu^2}{m_{Q}^2}\right)^{\epsilon}
\end{equation}
and,
\begin{equation}
 \begin{split}
\mathcal{F}^{1l}_{1,R} &= \cf \left(
\frac{1}{\epsilon} \Biggl[ - 1 + \biggl( 1 - \frac{1}{1-y} - \frac{1}{1+y} \biggr) H(0;y)\Biggr]  
- 2 - \left( \frac{1}{2}- \frac{1}{1-y} \right) H(0;y)
\right.\\
& \left.\qquad  - \biggl( 1 - \frac{1}{1-y} - \frac{1}{1+y} \biggr)  \bigl[4 \zeta(2)- 2 H(0;y) - H(0,0;y) - 2 H(1,0;y) \bigr]\right)\,,\\
\mathcal{F}^{2l}_{2,R} &= \cf \left( \left[ \frac{1}{1-y}-\frac{1}{1+y} 
\right] H(0;y) \right).\\
\end{split}
\end{equation}
We quote the one-loop renormalised vertex corrections 
for $\enen > 4 m^2$ 
given in \cite{bernreuther} up to order $\as$ and $\epsilon^0$ 
respectively. (Imaginary parts are omitted because they do not contribute 
to \order{\as}). The dimensionless variable $y$ is defined by 
$$ y = \frac{1-\sqrt{r_0}}{1+\sqrt{r_0}}.$$

The real emission contribution to the decay width $\gammareal$ is given by
\begin{equation}
\begin{split}
 \gammareal &= \frac{1}{2 \en} \left( \frac{1}{2} \sum_{spins,colours} \abs{\mtwo}^2\right) \Phi_2  (4 \pi \as \mu^{2 \epsilon}) \left(2 \cf \Aint\right)\,,\\
\end{split}
\label{eq:gammareal}
\end{equation}
while the averaged leading order matrix element squared proportional 
to $\abs{\mtwo}^2$ reads,
\begin{equation}
\sum_{spins,colours} \abs{\mtwo}^2 = (4 \pi \alpha Q_f^2) N_C \enen 2 \left( 3-r_0-2\epsilon \right)\,. 
\end{equation}
By expanding in $\eps$ our result for $\Aint$ given in terms 
of hypergeometric functions we could show that,
\begin{equation}
\gammareal + \gammavertex= \Gamma^{had}_{\gstar}
\end{equation}
giving us a strong check of our integrated massive final-final 
$Q\bar{Q}$ antenna $\Aint$.
\section{Conclusions and Outlook}
We have generalized the antenna subtraction method, originally developed for
massless final states to real radiation off massive final state 
partons at the next-to-leading order level.
The main building blocks of the antenna subtraction method are the 
antenna functions. Those encapsulate all unresolved radiation 
emitted between two colour-ordered hard radiators. 
As such those functions account for all unresolved radiation 
of the corresponding QCD matrix elements.
In Section 3, we have presented the final-final antenna functions 
for the following cases:
the presence of two massive final state radiators of equal masses or 
the presence of one massive and one massless radiator.
Section 4 contained a list of all non-vanishing unresolved limits of
the antennae functions presented in Section 3.  
Explicit phase space factorisation and parameterization formulae were 
presented in Section 5 where all massive final-final antennae 
functions were integrated over their corresponding phase space measure.

An important extension of subtraction methods is the combination with
parton shower algorithms \cite{Nagy,mcnlo,Weinzierl2,sherpaPS}, 
thus allowing for a full partonic event generation to NLO accuracy. 
So far, our massless final-final antenna functions are part of the parton
shower VINCIA \cite{vincia}.
With the formulation of the antenna subtraction
method for final state radiation off massive fermions presented here, 
it will become possible to construct 
antenna-based parton showers involving massive final state particles.

 \section*{Acknowledgements}
We wish to thank Thomas Gehrmann for useful discussions.
This research was supported by the Swiss National Science Foundation
(SNF) under contract PP002-118864 which is hereby acknowledged.

\bibliographystyle{JHEP}

\end{document}